# Elastic interactions of Fe adatoms on Cu(111) at the mesoscale

Wolfgang Kappus

Alumnus of ITP
Philosophenweg 19
D-69120 Heidelberg

wolfgang.kappus@t-online.de

## Abstract

An extended elastic eigenvector approach had earlier been developed to interpret ab-initio calculations of adatom interactions. It shows oscillating interactions as well as trio- and quarto (multisite) terms within clusters. It is now applied to the interaction of Fe adatoms on Cu(111).

The extended approach differs from previous calculations by using a sharp cutoff - generating oscillating interactions - and by taking into account interacting dimers - generating strong anisotropies and multisite terms. Additional weak anisotropy stems from the substrate and from the surface Brillouin zone shape.

This approach has 3 free parameters which have been fitted to first principles interactions of Fe adatoms on Cu(111). Elastic adatom pair interactions and dimer-dimer interactions at mesoscale separations show a reasonable good fit to the first principles interactions. At smaller separations elastic interaction values remain questionable. To enable a comparison with future first principle calculations also dimer-monomer interactions are shown.

Some conclusions on initial adatom formation and diffusion are proposed and open questions are formulated.

# 1. Introduction

Interactions of adatoms are a subject of continuous interest, various different interaction mechanisms have been described in detail [1].

Thanks to computational power and advanced algorithms the energies of adatom configurations have been determined from first principles using density-functional theory (DFT) [2] and references therein. Such ab-initio methods have been applied by Longo et al. to calculate interactions of Fe adatoms on Cu(111) surfaces in a mesoscopic range beyond the cut-off radius of electronic potentials [3]. Using electronic many-body potentials for computing, elastic interactions are claimed to be well justified. Single Fe adatom interactions here show an oscillating decay with the distance with an $s^{-3}$ amplitude. Strain calculations indicate the interactions elastic nature. The classical continuum theory of elastic interactions on surfaces [4, 5, 6] predicts an $s^{-3}$ tail but no oscillations. A review of elastic effects on surface physics was given in [7].

The interaction energies of Fe dimers calculated ab-initio on Cu(111) clearly indicate multisite effects. To interpret those effects an extension of the pair interaction theory of [5] to handle multisite terms is being used [8, 9].

In the pair interaction model of [5] the interaction of adatoms is mediated by their strain fields generated by single adatoms exerting isotropic stress to their vicinity. Isotropic stress is a consequence of adatom locations on sites with high symmetry. In a lattice description adatoms would exert forces to their immediate substrate neighbors creating a displacement field equivalent to a strain field.

In other cases where adatoms interact directly e.g. via their dipole moment or via electronic overlap, pairs of adatoms may create substrate strain by stretching or compressing the substrate to balance the forces. The strain field of such pairs will mediate interactions between pairs and also between pairs and monomers, in other words trio and quarto multisite interactions. If forces between dimer constituents are central, the model can be kept simple and the resulting stress field can (with symmetry restrictions) be described with one more free parameter [8] than the previous pair interaction model [5]. An oscillating interaction is the consequence of the sharp cutoff of a wave vector integral [9].

3 free parameters thus can be used to derive long range elastic interactions amongst adatoms and dimers fitting to DFT calculated interactions.

As Longo et al. pointed out [3] the detailed knowledge of adatom interactions allows also conclusions on adatom diffusion and on nucleation paths.

The paper is organized as follows:
After outlining the general motivation in section 1 and re-inspecting the DFT results of [3] in section 2, the interaction model is recalled in section 3. In section 4 elastic model parameters are fitted to the first principles calculations of [3]. In section 5 applicability and limitations of the model are discussed and open questions are addressed. Section 6 closes with a summary.

## 2. Re-inspection of DFT calculated interactions

In the following the adatom interactions calculated with DFT [3] will be re-inspected. They contain valuable information and are the base for the subsequent analysis. Their Fig. 1 shows pair interactions as function of distance and Fig. 4 shows dimer-dimer interactions as function of distance. The distances reflect the grid of high symmetry Fe adatom positions on Cu(111).

2.1. Surface grid

The high symmetry grid on the Cu(111) surface is spanned by the Cu crystal vectors  *a*={1,-1,0} and *b*={0,1,-1}. In the following surface coordinates *(m,n)* will be used, where *(1,0)*=*a* and *(0,1)*=*b*. In contrast to coordinates the basic data in Figs. 1 and 4 of [3] are labelled with surface directions like <100> and <010> which are not unique coordinates. It must be noted that *(1,0)* and *(0,1)* are not orthogonal but form an angle of 120°. Instead of coordinates also positions $s_i$ will be used below.

2.2. DFT data analysis

The DFT calculated interactions in Figs. 1 and 4 of [3] contain information valuable for the later parameter search. Pair interactions in Fig.1 of [3] obey an obvious $s^{-3}$ decay. Furthermore the most adatom-adatom pair interactions follow $V_{ad-ad}(s)=V_0*s^{-3}\cos(\kappa_{DFT}s)$ with $V_0=380$ meVÅ³ [3] and $\kappa_{DFT}=2\pi/s_0/\sqrt{2}=2.47$ Å⁻¹, as Fig1. below shows. Here $s_0$ denotes the lattice constant of copper. The interactions of [3] are marked as dots, the $V_{ad-ad}(s)$ dashed line is shown for comparison. Only DFT calculated interactions of adatoms with the distances 9.17Å, 11.66Å and 14.16Å fall beside and indicate deviations from isotropy. The DFT calculated interaction at 7.63Å is slightly below the $s^{-3}$ interpolation which indicates the limit of the elastic interpretation below 6Å.

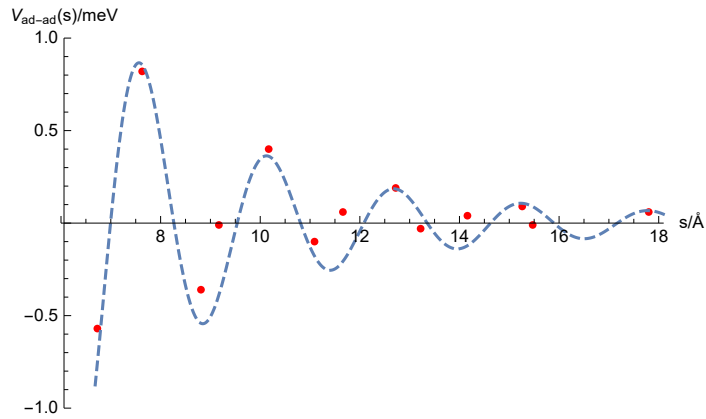

Figure 1. Comparison of pair interactions as function of distance taken from Fig.1 of [3] (red dots) and $V_{ad-ad}(s) = V_0*s^{-3}\cos(\kappa_{DFT}s)$ (dashed line).

The interaction energy of two dimers is the sum of adatom pair- and multisite interactions. As dimer two adjacent adatoms are denoted. Multisite lateral interactions are trio- or quarto interactions of 4 adatom clusters. Comparison of adatom pair interactions from Fig. 1 with dimer-dimer interactions from Fig. 4, both of [3] thus allows to identify multisite contributions. Electronic interactions between the two 2nd nearest neighbor constituents of a dimer are not considered in this mesoscale analysis.

Table 1 shows the short configuration, the vector between dimer centers, in column1. In column 2 the distance between the dimer centers are listed. Column 3 lists the coordinates of the 4 adatoms forming 2 dimers. In column 4 the sum of pair interactions of the 4 adatoms taken from Fig.1 of [3] are listed. column 5 shows the dimer-dimer interactions of this distance taken from Fig.4 of [3].

| short config. | distance/Å | dimer config. | pair int.sum/meV | dim.int./meV |
|---|---|---|---|---|
| $\begin{pmatrix} 3 \\ 1 \end{pmatrix}$ | 6.73 | $\begin{pmatrix} 0 & 1 & 3 & 4 \\ 0 & 2 & 1 & 3 \end{pmatrix}$ | −1.72 | 0.63 |
| $\begin{pmatrix} 3 \\ 0 \end{pmatrix}$ | 7.63 | $\begin{pmatrix} 0 & 1 & 3 & 4 \\ 0 & 2 & 0 & 2 \end{pmatrix}$ | 0.92 | 1.72 |
| $\begin{pmatrix} 4 \\ 2 \end{pmatrix}$ | 8.81 | $\begin{pmatrix} 0 & 1 & 4 & 5 \\ 0 & 2 & 2 & 4 \end{pmatrix}$ | 0.16 | 0.62 |
| $\begin{pmatrix} 4 \\ 3 \end{pmatrix}$ | 9.17 | $\begin{pmatrix} 0 & 1 & 4 & 5 \\ 0 & 2 & 3 & 5 \end{pmatrix}$ | −0.4 | 0.87 |
| $\begin{pmatrix} 4 \\ 0 \end{pmatrix}$ | 10.17 | $\begin{pmatrix} 0 & 1 & 4 & 5 \\ 0 & 2 & 0 & 2 \end{pmatrix}$ | 0.6 | 1.06 |
| $\begin{pmatrix} 5 \\ 3 \end{pmatrix}$ | 11.09 | $\begin{pmatrix} 0 & 1 & 5 & 6 \\ 0 & 2 & 3 & 5 \end{pmatrix}$ | −0.17 | 0.51 |
| $\begin{pmatrix} 5 \\ 4 \end{pmatrix}$ | 11.66 | $\begin{pmatrix} 0 & 1 & 5 & 6 \\ 0 & 2 & 4 & 6 \end{pmatrix}$ | −0.15 | 0.6 |
| $\begin{pmatrix} 6 \\ 5 \end{pmatrix}$ | 14.16 | $\begin{pmatrix} 0 & 1 & 6 & 7 \\ 0 & 2 & 5 & 7 \end{pmatrix}$ | 0.04 | 0.37 |
| $\begin{pmatrix} 7 \\ 3 \end{pmatrix}$ | 15.47 | $\begin{pmatrix} 0 & 1 & 7 & 8 \\ 0 & 2 & 3 & 5 \end{pmatrix}$ | 0.08 | 0.28 |

Table 1. Comparison of full dimer-dimer interactions with the pair interaction part of dimer configurations. First column: short configuration. Second column: distance between two dimers. Third column: full configuration of two dimers. Forth column: sum of pair interactions taken from Fig. 1 of [3]. Fifth column: dimer-dimer interaction taken from Fig. 4 of [3].

When comparing column 4 and 5 of Table 1 caution is necessary because the interaction of dimer-dimer configurations with identical center distance can differ significantly. For example short configurations (4,2) and (2,4) have the same distance but are quite different. The elastic dimer-dimer repulsion for the (2,4) configuration is 7.56 meV due to a strong multisite effect - almost compensated by the short range electronic interaction of the (1,2)-(2,4) link. The only unique configurations are (n,0).

When restricting to (n,0) configurations the comparison of column 4 and 5 of Table 1 indicates that multisite effects are relevant and can lead to increased repulsion of dimers. The multisite repulsion part of e.g. the (3,0) configuration is 0.8 meV, comparable in size to the pair alone part.

Some narrow configurations like (3,2) do not appear in Table 1; such dimer configurations contain adatom pairs with (2,0) or 5.09 Å distance, which have not been calculated in [3].

The elastic nature of the Fe-Cu(111) interactions on the mesoscale is confirmed by the inset in Fig. 4 of [3].

# 3. Adatom Interaction Model

In this section the elastic eigenvector model used is recalled from [8,9]. For convenience the concept of the monomer and dimer stress fields is repeated and the elastic eigenvalues for Cu(111) surfaces are given. The interaction equation for the different cases is shown and the influence of the Brillouin zone shape on the interaction is discussed.

3.1. Elastic interaction model

While elastic pair interactions are related to the stress fields of single adatoms, dimers will generate multisite interactions.

Adatoms or dimers exert forces on their substrate neighbors leading to a displacement of substrate atoms to balance those forces. Such displacements will increase or decrease the energy of neighboring adatoms or dimers. In a continuum description adatoms and dimers exert stress parallel to the surface leading to substrate strain which in turn can lead to an attraction or a repulsion of neighboring adatoms or dimers. The strength of such strain mediated interaction depends on the stress adatoms or dimers exert on the substrate and on the stiffness of the substrate.

The elastic energy of a substrate with adatoms in a continuous description is given by the sum of two parts, the energy of the distorted substrate and the energy of adatoms exerting tangential forces on the substrate

$$H_{el} = \frac{1}{2} \int_V \epsilon(r) \, c \, \epsilon(r) \, dr + \int_S \epsilon(s) \, \pi(s) \, ds \,. \tag{3.1}$$

Here $\epsilon = [\epsilon_{\alpha\beta}]$ denotes the strain tensor field, $c = [c_{\alpha\beta\mu\nu}]$ denotes the elastic constants tensor, and $\pi = [\pi_{\mu\nu}]$ denotes the force dipole- or stress tensor field. The integrals comprise the bulk V or the surface S. The strain field $\epsilon(r)$ is related to the displacement field $u(r)$ by

$$\epsilon_{\alpha\beta}(r) = \frac{1}{2} \left( \nabla_\alpha u_\beta(r) + \nabla_\beta u_\alpha(r) \right) \,. \tag{3.2}$$

Following [8] the stress field $\pi(s)$ is superimposed of $n_t$ different types $\pi_k(s)$

$$\pi(s) = \sum_{k=1}^{n_t} \pi_k(s) = \sum_{k=1}^{n_t} P_k \rho_k(s) \,, \tag{3.3}$$

where we have introduced $n_t$ different types of stress tensors $P_k$ and of adatom monomer or dimer densities $\rho_k(s)$.

On (111) surfaces with adatom positions on high symmetry sites $n_t=4$ i.e. $k$ can take the values 1 to 4 where

$P_1 = P_1[\delta_{\alpha\beta}]$ stand for an isotropic monomer stress tensor (or force dipole tensor),

$P_2 = P_2[\delta_{\alpha 1} \delta_{\beta 1}]$ stands for an anisotropic dimer stress tensor. $P_3$ and $P_4$ are the 120° / 240° rotated equivalents of $P_2$. On (111) surfaces with adatom positions on high symmetric sites $P_4 = P_3 = P_2$.

$\rho_1(s)$ then stands for the adatom monomer density distribution, $\rho_2(s)$ for the x-directed adatom dimer density distribution and $\rho_3(s)$, $\rho_4(s)$ for the rotated equivalents of $\rho_2(s)$.

With Eq. (3.3) Eq. (3.1) reads

$$H_{el} = \frac{1}{2} \int_V \epsilon(r) \, c \, \epsilon(r) \, dr + \sum_{k=1}^{n_t} \int_S \epsilon(s) \, P_k \, \rho_k(s) \, ds \,. \tag{3.4}$$

The strain field $\epsilon(r)$ is determined for given densities $\rho_k(s)$ by the requirement of mechanical equilibrium

$$\delta H_{el} / \delta u_\alpha(r) = 0 \,. \tag{3.5}$$

After an expansion in plane waves and calculations beyond the scope of this paper we end up at [8]

$$H_{el} = \frac{1}{2} \sum_{k=1}^{n_t} \sum_{l=1}^{n_t} \int_S \int_S \rho_k(s) \, V_{kl}(s-s') \, \rho_l(s') \, ds \, ds', \tag{3.6}$$

with the elastic interaction $V_{kl}$ between adatoms (or dimers) of type $k$ at $0$ and adatoms (or dimers) of type $l$ at $s = (s, \chi)$

$$V_{kl}(s-s') = (2\pi)^{-1} \sum_p \omega_{kl,p} \cos(p\chi) \cos(p\pi/2) \int_0^{\kappa_{BZ}} \kappa^2 J_p(\kappa s) \, d\kappa \,, \tag{3.7}$$

where $J_p(\kappa s)$ denotes the Bessel function of order $p$ and $\omega_{kl,p}$ denotes eigenvalues to be discussed below. The $\kappa$ integral with a hard limit at the surface Brillouin zone $\kappa_{BZ}$ results in a generalized hypergeometric function which together with its coefficients decays oscillating like $s^{-3/2}$. In the sections below the interaction

$$V_{kl}(s, \chi) = (2\pi)^{-1} \sum_p \omega_{kl,p} \cos(p\chi) \cos(p\pi/2) \, 2^{-1-p} \kappa_{BZ}^3 \, (s\kappa_{BZ})^p \, \Gamma\left(\frac{3+p}{2}\right) *$$

$$_1F_2\left(\left(\frac{3+p}{2}\right), \left(\frac{5+p}{2}, 1+p\right), -\frac{1}{4} s^2 \kappa_{BZ}^2\right) X(s), \qquad (3.8)$$

with the heuristic factor

$$X(s) = (s/s_0)^{3/2} / \left(2 + (s/s_0)^3\right) \qquad (3.9)$$

is used, ensuring proper mesoscale $s^{-3}$ decay and a regular behavior for $s<6$ Å where the simple elastic approach does not hold.

In (3.8) $\Gamma(p)$ denotes the Gamma function and $_1F_2(a,b_1,b_2,s)$ the generalized hypergeometric function. $_1F_2$ is the appropriate name for the $_pF_q$ function with one factor $a$ and two factors $b_i$ in Eq. (3.8). $\kappa_{BZ}$ is the wave vector distance between the origin and the surface Brillouin zone in $\chi$ direction. $\omega_{kl,p}$ are eigenvalues for the 4x4 interaction types $(k,l)$ with $p$ resulting from a series expansion. $p$ takes the values 0, 6, 12 in the monomer case and 0, 2, 4 in the dimer case. The eigenvalues $\omega_{kl,p}$ are proportional to the product of stress parameters $P_k P_l$ and inversely proportional to the elastic constant $c_{44}$ defining a dimensionless constant $\hat{\omega}_{kl,p}$

$$\omega_{kl,p} = \hat{\omega}_{kl,p} P_k P_l / c_{44} . \qquad (3.10)$$

Table 2 shows the dimensionless coefficients $\hat{\omega}_{kl,p}$ for Cu(111) using elastic constants $c_{11}=169$, $c_{12}=122$, $c_{44}=75.3$ GPa [10].

| k | l | $\hat{\omega}_{kl,0}$ | $\hat{\omega}_{kl,6}$ | $\hat{\omega}_{kl,12}$ |
|---|---|---|---|---|
| 1 | 1 | −1.0096 | −0.0031 | −0.0042 |

| k | l | $\hat{\omega}_{kl,0}$ | $\hat{\omega}_{kl,2}$ | $\hat{\omega}_{kl,4}$ |
|---|---|---|---|---|
| 1 | 2 | −0.716 | 0.719 | −0.003 |
| 2 | 2 | −1.2 | −1.017 | 0.183 |
| 3 | 4 | −0.157 | −0.492 | 0.192 |

Table 2. Coefficients $\hat{\omega}_{kl,p}$ for Cu(111) calculated with the elastic eigenvector approach of [8].

The $k=l=1$ coefficients belong to a monomer-monomer (pair) interaction; the small $\hat{\omega}_{11,6}$ and $\hat{\omega}_{11,12}$ values reflect the small elastic anisotropy of the substrate's copper (111) surface in contrast to the following multisite coefficients showing strong anisotropy.

The $k=1$, $l=2$ coefficients belong to monomer-dimer (trio) interactions.

The $k=2$, $l=2$ coefficients belong to (quarto) interactions of two parallel dimers.

The $k=3$, $l=4$ coefficients belong to (quarto) interactions of two dimers spanning an angle of 120°.

$X(s)$ in Eq. (3.9) differs from a similar factor in Eq. (2.14) of [9]. $X(s)$ fits better to the DFT data base of Fe-Cu(111) when sub-mesoscale distances are involved; the data base of Fe-Cu(111) [3] is more complete than the data base of O-Pd(100) used in [9].

$V_{kl}(s, \chi)$ in Eq. (3.8) shows features helpful for interpreting of the DFT interaction data:

It looks similar to $V_0 * s^{-3} \cos(\kappa_{BZ} s)$ for $s>2$ (cf. section 2.2). $V_{kl}(s, \chi)$ is regular at the origin and its wavelength is $\kappa_{BZ}$.

For $\kappa_{BZ}=2\pi/s_0$ it has roots near $s=s_0 n/2$, with $n=2,3,4,...$ and the $p>0$ terms converge with $s^{-3}$ like the $p=0$ term.

# 4. Parameter search and calculation results

In the following section the parameters for the eigenvector model will be searched on the base of the DFT data of Longo et al. [3]. The results will allow checking if an interpretation of those data by the elastic eigenvector model is valid.

A detailed look on Fig.4 of [3] shows that Fe dimer constituents are 2nd nearest neighbors on the grid, indicating Fe adatoms are larger than Cu bulk atoms. Those Fe dimers are assumed to consist of overlapping Fe atoms. Their chemical bonds together with the bonds to adjacent bulk atoms would create the $P_2$ related stress. The details of the chemical bonds - fortunately - are not relevant for the elastic model.

Eq. (3.8) is the result of integration over the surface Brillouin zone in Eq. (3.7), the shape of which dictates the exact value of $\kappa_{BZ}(\chi)$. In the following section the search for the appropriate shape will be discussed. Also the question will be answered if the effective 2 parameter set of the elastic model need to be extended.

### 4.1. Hexagonal and similar surface Brillouin zones

The shape of the surface Brillouin zone affects the interaction values of Eq. (3.8) via $\kappa_{BZ}$. $\kappa_{BZ}(\chi)$ is the wave vector distance between the origin and the surface Brillouin zone in $\chi$ direction. The regular surface Brillouin zone is a hexagon but the fit results using this zone are less convincing. Since shapes similar to a hexagon can improve the fit, an optimum shape is searched. The amount of interaction data given by [3] allows trying shapes ranging from the correct hexagon to a circle. It turned out that a circle deformed towards a hexagon

$$\kappa_{BZ}(\chi) = \kappa_{BZ0}\left(1 - z_{BZ} * \sin(3\chi)^6\right) . \qquad (4.1)$$

fits best. This shape reflects the symmetry of the (111) surface. The exponent 6 creates a small peak like deformation of a circle.

Critical adatom distances for parameter fits are at the roots of $V_{kl}(s, \chi)$ in Eq. (3.8); for $\kappa_{BZ}=2\pi/s_0/2^{1/2}$ they are near 7.9, 9.2, 10.5, 11.7, 13, 14.3, 15.6 Å. Small distance changes near the roots lead to significant interaction changes. Some of the root values appear also in the discussion of section 2.2.

### 4.2. Parameter search

The stress parameters $P_1$ and $P_2$ together with the surface Brillouin zone deformation parameter $z_{BZ}$ are the unknown model parameters in Eq. (3.8) and will be determined by a fit to the DFT calculated interactions. Instead of running a simultaneous optimization the following steps have been performed:
In the first step the circular surface Brillouin zone limit $\kappa_{BZ0}$ is assumed $2\pi/s_0/\sqrt{2}$, where $s_0$ denotes the Cu bulk lattice constant. This is the "natural" choice for the (111) surface.
In the second step the stress parameter $P_1$ and the deformation parameter $z_{BZ}$ (defined in Eq. (4.1)) are determined by fitting the pair interaction $V_{11}(s, \chi)$ to prominent $V(m,n)$ values in Fig.1 of [3].
In the last step the stress parameter $P_2$ is then determined by fitting the sum of multisite interactions and pair interactions to the interaction values $V(s_1, s_2, s_3, s_4)$ of selected dimers on positions $s_1, s_2$ and $s_3, s_4$ in Fig.4 of [3].

### 4.3. Model parameters fitted

The parameter values of the elastic model optimized for fitting to the DFT calculated pair interactions in Fig. 1 and dimer-dimer interactions in Fig. 4 of [3] are the set {$P_1$=1.76 meV, $P_2$=1.00 meV, and $z_{BZ}$=0.085} together with the constant $\kappa_{BZ0}=2\pi/s_0/\sqrt{2}$.
The numerical values for $P_1$ and $P_2$ were derived under the convenience assumption $c_{44}s_0^3$=1 meV for allowing a direct evaluation of Eq. (3.8).
The (3,2) and (3,4) configurations contain short range (2,0) and (2,2) links; their repulsive interaction accord-ing to Eq. (3.8) is far below the $s^{-3}$ envelope. As noted in section 3.1, Eq. (3.9) is a heuristic approximation for handling adatom distances $s<6$Å.

The variance of the model parameters $P_1$, $P_2$ and $z_{BZ}$ is below 10%.

### 4.4. Adatom-adatom interaction results

In this section the interactions according to Eq. (3.8) with the model parameters shown in section 4.3 are presented.

Fig.2 shows the elastic pair interactions of Fe adatoms as function of their distance as points connected by dotted lines. For comparison the results in Fig. 1 of [3] are included as red points. The oscillations are clearly reproduced but a few pairs like (5,1), 11.66 Å apart and (6,1), 14.16 Å apart disturb the picture. They belong to the critical distances discussed in section 4.1.

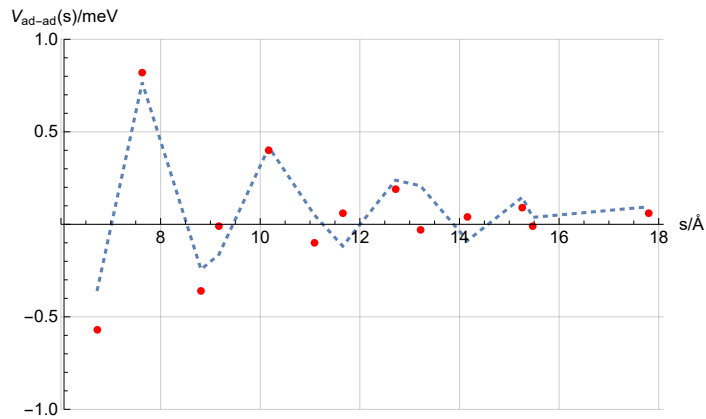

Figure 2. Comparison of pair interactions $V_{ad-ad}$ as a function of distance $s$ taken from Fig.1 of [3] (red dots) with $V_{kl}(s, \chi)$ values calculated with Eq. (3.8) (connected by dotted lines).

### 4.5. Dimer-dimer interaction results

Fig.3 shows the elastic interactions of Fe dimers as function of their distance as points connected by dotted lines. For comparison the results in Fig. 4 of [3] are included as red dots. The interactions here are repulsive and the fit is reasonable except for the dimer distances 9.17, 11.66, 14.16, 15.47, 17.8 Å.

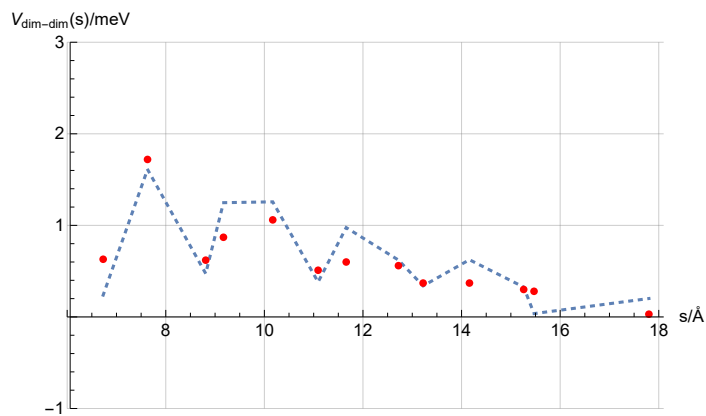

Figure 3. Comparison of dimer-dimer interactions $V_{dim-dim}$ as function of distance $s$ taken from Fig.4 of [3] (red dots) with the dimer-dimer interaction values calculated with Eq. (3.8) (connected by dotted lines).

Tab. 3 shows the dimer-dimer interaction results in numerical form in column 5; for completeness the pair only interaction part is listed in column 4 as well as the DFT results taken from Fig.4 of [3] in column 6. Major deviations between elastic and DFT interactions occur at the configurations (3,2), caused by the (2,0) extreme short range interaction, at (3,4), caused by the (2,2) extreme short range interaction and at (4,5), (7,3), (5,6) near the root distances.

| short config. | distance/Å | dim.–dim. config. | pair only int./meV | dim.–dim. int./meV | DFT dim.–dim. int. /meV |
|---|---|---|---|---|---|
| $\begin{pmatrix}3\\2\end{pmatrix}$ | 6.73 | $\begin{pmatrix}0&1&3&4\\0&2&2&4\end{pmatrix}$ | 1.06 | 0.24 | 0.63 |
| $\begin{pmatrix}3\\0\end{pmatrix}$ | 7.63 | $\begin{pmatrix}0&1&3&4\\0&2&0&2\end{pmatrix}$ | 1.04 | 1.61 | 1.72 |
| $\begin{pmatrix}4\\2\end{pmatrix}$ | 8.81 | $\begin{pmatrix}0&1&4&5\\0&2&2&4\end{pmatrix}$ | 0.16 | 0.47 | 0.62 |
| $\begin{pmatrix}3\\4\end{pmatrix}$ | 9.17 | $\begin{pmatrix}0&1&3&4\\0&2&4&6\end{pmatrix}$ | 1.00 | 1.25 | 0.87 |
| $\begin{pmatrix}4\\0\end{pmatrix}$ | 10.17 | $\begin{pmatrix}0&1&4&5\\0&2&0&2\end{pmatrix}$ | 0.93 | 1.26 | 1.06 |
| $\begin{pmatrix}5\\2\end{pmatrix}$ | 11.09 | $\begin{pmatrix}0&1&5&6\\0&2&2&4\end{pmatrix}$ | 0.50 | 0.38 | 0.51 |
| $\begin{pmatrix}4\\5\end{pmatrix}$ | 11.66 | $\begin{pmatrix}0&1&4&5\\0&2&5&7\end{pmatrix}$ | 0.53 | 0.98 | 0.6 |
| $\begin{pmatrix}5\\0\end{pmatrix}$ | 12.72 | $\begin{pmatrix}0&1&5&6\\0&2&0&2\end{pmatrix}$ | 0.44 | 0.62 | 0.56 |
| $\begin{pmatrix}6\\3\end{pmatrix}$ | 13.22 | $\begin{pmatrix}0&1&6&7\\0&2&3&5\end{pmatrix}$ | 0.30 | 0.34 | 0.37 |
| $\begin{pmatrix}5\\6\end{pmatrix}$ | 14.16 | $\begin{pmatrix}0&1&5&6\\0&2&6&8\end{pmatrix}$ | 0.42 | 0.62 | 0.37 |
| $\begin{pmatrix}6\\0\end{pmatrix}$ | 15.26 | $\begin{pmatrix}0&1&6&7\\0&2&0&2\end{pmatrix}$ | 0.22 | 0.33 | 0.3 |
| $\begin{pmatrix}7\\3\end{pmatrix}$ | 15.47 | $\begin{pmatrix}0&1&7&8\\0&2&3&5\end{pmatrix}$ | −0.01 | 0.04 | 0.28 |
| $\begin{pmatrix}7\\0\end{pmatrix}$ | 17.8 | $\begin{pmatrix}0&1&7&8\\0&2&0&2\end{pmatrix}$ | 0.13 | 0.20 | 0.03 |

Table 3. Interactions of dimer-dimer configurations calculated with the elastic eigenvector model Eq. (3.8), compared with ab initio calculated. First column: short configuration. Second column: distance between two dimers. Third column: full configuration of two dimers. Forth column: sum of pair interactions. Fifth column: dimer-dimer interaction calculated with Eq. (3.8). Sixth column: DFT calculated dimer-dimer interactions taken from Fig.4 of [3].

The differences between pair only interactions (column 4) and interactions including multisite effects (column 5) are significant, especially for the short range configurations. Also the (n,0) dimer configurations show considerable additional repulsion by multisite effects.

4.6. Dimer-adatom interaction results

To enable more detailed comparisons with future DFT calculations trio interactions are shown in Tab.4. In column 5 the full elastic interactions are listed for various trio configurations while in column 4 the pair only part is given. The differences are less pronounced, except for the (3,4) configuration with the short range (2,2) link and the (4,5) configuration where the adatom is located closely to the dimer direction.

| short config. | distance/Å | dim.–at. config. | pair only int./meV | dim.–at. int./meV |
|---|---|---|---|---|
| $\begin{pmatrix}3\\2\end{pmatrix}$ | 5.54 | $\begin{pmatrix}0 & 1 & 3\\0 & 2 & 2\end{pmatrix}$ | 0.99 | 1.12 |
| $\begin{pmatrix}3\\0\end{pmatrix}$ | 7.94 | $\begin{pmatrix}0 & 1 & 3\\0 & 2 & 0\end{pmatrix}$ | 0.52 | 0.56 |
| $\begin{pmatrix}4\\2\end{pmatrix}$ | 7.94 | $\begin{pmatrix}0 & 1 & 4\\0 & 2 & 2\end{pmatrix}$ | 0.52 | 0.56 |
| $\begin{pmatrix}3\\4\end{pmatrix}$ | 7.08 | $\begin{pmatrix}0 & 1 & 3\\0 & 2 & 4\end{pmatrix}$ | 1.18 | 2.05 |
| $\begin{pmatrix}4\\0\end{pmatrix}$ | 10.40 | $\begin{pmatrix}0 & 1 & 4\\0 & 2 & 0\end{pmatrix}$ | 0.47 | 0.48 |
| $\begin{pmatrix}5\\2\end{pmatrix}$ | 10.40 | $\begin{pmatrix}0 & 1 & 5\\0 & 2 & 2\end{pmatrix}$ | 0.47 | 0.48 |
| $\begin{pmatrix}4\\5\end{pmatrix}$ | 9.60 | $\begin{pmatrix}0 & 1 & 4\\0 & 2 & 5\end{pmatrix}$ | 0.65 | 1.08 |
| $\begin{pmatrix}5\\0\end{pmatrix}$ | 12.90 | $\begin{pmatrix}0 & 1 & 5\\0 & 2 & 0\end{pmatrix}$ | 0.22 | 0.23 |
| $\begin{pmatrix}6\\3\end{pmatrix}$ | 12.30 | $\begin{pmatrix}0 & 1 & 6\\0 & 2 & 3\end{pmatrix}$ | 0.09 | 0.08 |
| $\begin{pmatrix}5\\6\end{pmatrix}$ | 12.10 | $\begin{pmatrix}0 & 1 & 5\\0 & 2 & 6\end{pmatrix}$ | 0.42 | 0.42 |
| $\begin{pmatrix}6\\0\end{pmatrix}$ | 15.40 | $\begin{pmatrix}0 & 1 & 6\\0 & 2 & 0\end{pmatrix}$ | 0.11 | 0.11 |
| $\begin{pmatrix}7\\3\end{pmatrix}$ | 14.70 | $\begin{pmatrix}0 & 1 & 7\\0 & 2 & 3\end{pmatrix}$ | −0.05 | −0.05 |
| $\begin{pmatrix}7\\0\end{pmatrix}$ | 17.90 | $\begin{pmatrix}0 & 1 & 7\\0 & 2 & 0\end{pmatrix}$ | 0.07 | 0.07 |

Table 4. Interactions of dimer-adatom configurations calculated with the elastic eigenvector model Eq. (3.8). First column: short configuration. Second column: distance between a dimer and an adatom. Third column: full configuration of a dimer and an adatom. Forth column: sum of pair interactions. Fifth column: full dimer-adatom interaction.

### 4.7. Preferred aggregation paths

Though the elastic adatom-adatom interactions are in the meV range only, some conclusions on preferred paths for aggregation can be drawn. While (n,0) configurations are repulsive, the paths (4,1) or (4,2) or (4,3) -> (3,1) or (3,2) -> (2,1) for approaching an (0,0) adatom would be preferred. The (2,1) position is a stable dimer with electronic interactions dominating.

When a dimer has formed, the elastic interactions for catching a further adatom or dimer are almost repulsive according to Tables 3 and 4. The existence of further dimer positions with attractive interactions, however, must be noted. They are not shown as in [3].

When discussing adatom- or dimer diffusion one has to keep in mind that the associated strain cloud reduces their mobility, a quasiparticle effect.

# 5. Discussion

5.1. Restrictions and limitations of the model

The DFT calculations of Longo et al. and their conclusions were restricted to the mesoscale [3]. At very small distances electronic interactions dominate, typically decaying with $s^{-5}$; at larger distances surface-state electronic interactions dominate decaying with $s^{-2}$ [11]. In the intermediate range elastic interactions prevail at least for the Fe-Cu(111) system [3]. The mesoscale restriction is also valid for the present continuous elastic model; it cannot be expected to cover very small distances.

The small distance limitations of the elastic model are reasons for apparent shortcomings concerning the interpretation of certain first principles dimer-dimer interactions of [3], because those configurations contain sub-mesoscale distances.

Further shortcomings are related with the surface Brillouin zone shape and the $\kappa$ integration cutoff. The shape is not the natural choice and the sharp cutoff leads to a function which does not follow the classical $s^{-3}$ law though it is oscillating.

The model is based on the theory of elasticity in the substrate and on the lateral stress adatoms apply to the surface. Key assumption is the mechanism by which adatoms and dimers interact. Monomer adatoms sitting on high symmetry adatom sites expand or contract the substrate by creating isotropic stress. Multisite interactions are due to the anisotropic stress adatom pairs exert to the substrate when they are bound electronically and stretched (or compressed) due to their position on substrate sites.

The restriction to high symmetry adatom locations has the advantage of stress parameter $P_k$ degeneracy. On (111) surfaces $P_4 = P_3 = P_2$ due to the equivalence of dimers with a 120° angle. This reduces the number of free model parameters.

A further key assumption is an ideal flat surface, i.e. the absence of steps.

The present model parameters rely on the DFT calculations in [3]. The adatom configuration base and their DFT energies used in this paper are restricted to the configurations published; they are are not complete but allow a reasonable fit to the elastic eigenvector model.

Multisite effects [12] are clearly visible but not apparent in all details.

5.2. Open questions and further aspects

The search for a cutoff function in Eq. (3.7) between a smooth exponential Exp($-\kappa^2 s^2$) and a hard Heaviside function $\theta(\kappa\text{-}s)$ could lead to an improved oscillating interaction with a proper decay and phase. As shown in [8] the smooth exponential cutoff leads to elastic interactions with an $s^{-3}$ decay without oscillation.

DFT calculations of short distant adatoms could improve the basis for better fits; especially the (2,0) adatom-adatom configuration would help because some dimer-dimer configurations contain such small distance links. The minimum distances provided in [3] are (3,1) and (3,0). Most interesting would be DFT calculations of the (1,2) nearest neighbor configuration to cover also the electronic interaction part.

There are a number of possible orientations of the Fe-Fe dimer with respect to the Cu (111) surface, both with respect to angle (to the Cu (111) crystallographic axis) as well as with respect to site (atop, versus hollow) and a wide range of possible placement of the image plane. In addition to the placement of any dimer, on Cu (111), there is also the issue of the relative placement and orientation of one dimer with respect to another. Expanding the range of possible dimer to dimer placements, in various sites, would add insight into the nature of the dimer to dimer interaction and help in the identification of other attractive parallel dimer-dimer configurations. Possible routes to this include machine learning, in the context of density function theory. But to compare with experiment, thermal effects, no doubt, should be included, suggesting that DMFT-DFT might also be valuable for further insight.

Especially interesting for the search of attractive parallel dimer-dimer configurations are the (n,1) dimer positions because the (n,1) adatom positions are attractive. An example for non parallel dimer orientations is the {(0,0),(1,2),(2,0),(4,1)} configuration; in this case also the $V_{13}$ tri- and $V_{23}$ quarto interactions would contribute in addition to $V_{11}$, $V_{12}$, $V_{22}$ in the parallel dimer orientation case.

A theoretical (DFT) model to determine the magnitude of the stress parameters $P_k$ could determine the elastic adatom interaction ab initio. For this purpose the lateral displacements of adjacent substrate atoms an adatom creates would have to be calculated as well as the lateral forces between adatom and those substrate atoms.

The trio values of Tab. 4 provide a test challenge for the present model to be compared with future DFT calculations. While in dimer-dimer configurations pair-, trio- and quarto interactions contribute, in trio adatom configurations only pair- and trio interactions are acting. Such test would help to verify the elastic model.

Other systems could be interesting to be analyzed and interpreted with the elastic eigenvector method. Systems with e.g. first and second nearest neighbor repulsion and third and fourth nearest neighbor attraction like O-Pd(100) may show mesoscale oscillating interactions. (111) surfaces are especially well suited for finding oscillations because elastic anisotropies are smaller than on (100) surfaces.

STM experiments to identify oscillating substrate displacements caused by adatoms could be interesting. But distinguishing them from Friedel oscillations seems to be an issue.

# 6. Summary

The interaction energies of Fe adatoms and dimers on Cu(111), calculated ab initio [3], have been interpreted by an elastic eigenvector model with 3 parameters. The interpretation fits the basic findings of [3], especially the oscillating $s^{-3}$ interactions of adatoms and multisite effects when dimers interact. Deviations are explained by short range- and Brillouin zone effects as expected from a continuous elastic model. Ideas to broaden the DFT data base and to search for an improved interaction equation are sketched.

# Acknowledgement

This work was initiated by the unknown referee of [9]. Many thanks for his hint on ref. [3]. Many thanks also to the current referees for their substantial comments.